\documentclass[reprint, prl, nofootinbib, superscriptaddress]{revtex4-1}

\usepackage[utf8]{inputenc}
\usepackage{bm, amsmath, amsfonts, amssymb, color, graphicx, hyperref, natbib}
\hypersetup{colorlinks=true, linkcolor=blue, filecolor=blue, urlcolor=blue, citecolor=blue}

\begin{document}

\title{Detectability of Gravitational Waves from the Coalescence of Massive \\ Primordial Black Holes with Initial Clustering}

\author{Qianhang Ding}
\affiliation{Department of Physics, The Hong Kong University of Science and Technology, Hong Kong, P.R.China}
\affiliation{Jockey Club Institute for Advanced Study, The Hong Kong University of Science and Technology, Hong Kong, P.R. China}

\author{Tomohiro Nakama}
\affiliation{Jockey Club Institute for Advanced Study, The Hong Kong University of Science and Technology, Hong Kong, P.R. China}

\author{Joseph Silk}
\affiliation{Sorbonne Universit\`es, UPMC Univ Paris 6 et CNRS, UMR 7095, Institut d'Astrophysique de Paris,
98 bis bd Arago, F-75014 Paris, France}
\affiliation{Department of Physics and Astronomy, The Johns Hopkins University, Baltimore, MD 21218, USA}
\affiliation{BIPAC, University of Oxford,1 Keble Road, Oxford OX1 3RH, UK}

\author{Yi Wang}
\affiliation{Department of Physics, The Hong Kong University of Science and Technology, Hong Kong, P.R.China}
\affiliation{Jockey Club Institute for Advanced Study, The Hong Kong University of Science and Technology, Hong Kong, P.R. China}

\begin{abstract}
    We show that the effect of initial  non-Gaussian clustering can significantly enhance the event rate for primordial black hole (PBH) coalescence. The impact of such clustering is studied in a specific scenario of multi-stream inflation. Initial clustering enables the possibility of detecting massive PBH coalescence by space-based gravitational wave interferometers such as LISA and DECIGO/BBO. The parameter regime for the ground-based detectors to detect PBH coalescence is also  extended. 
\end{abstract}

\maketitle

Black holes of a variety of masses could have formed in the early Universe. They are known as primordial black holes (PBHs) \cite{Hawking:1971ei,Dovich:1966no,Carr:1974nx}. PBHs with initial mass less than $10^{15}$g have already evaporated by now through Hawking radiation \cite{Hawking:1974rv}. More massive PBHs, once formed, are still present in the Universe today, constituting part of the dark matter. See \cite{Carr:2009jm,Carr:2016drx,Sasaki:2018dmp} for production mechanisms and observational constraints on PBHs of different masses. 

Recently, stellar mass PBHs have received renewed interest \cite{Bird:2016dcv,Clesse:2016vqa,Sasaki:2016jop} since the first detection of gravitational waves from the coalescence of a black hole binary  \cite{Abbott:2016blz}. Stellar-mass PBHs could have formed sufficiently frequently in the early Universe so that mergers of PBHs may be observable by ongoing and/or planned ground-based gravitational-wave interferometers, such as LIGO \cite{Abramovici:1992ah}, Virgo \cite{TheVirgo:2014hva} and KAGRA \cite{Somiya:2011np}. 

Targeting much lower frequencies, future space-based gravitational-wave interferometer experiments are planned, such as LISA \cite{LISA}, DECIGO \cite{Kawamura:2006up}, Taiji \cite{Hu:2017mde} and Tianqin \cite{Luo:2015ght}. It thus becomes interesting to consider the possibility of observing mergers of PBHs with heavier masses. 

In the simplest PBH formation scenarios, mergers of massive PBHs in the sensitive frequency ranges of space-based interferometers are unlikely to be detectable, due to  tight constraints on the PBH abundance from various astrophysical observations. This can be seen by noting that, in \cite{Sasaki:2016jop}, assuming a uniform distribution of PBHs
\footnote{Even for PBHs which are formed from Gaussian primordial fluctuations, strictly speaking, this assumption is over-simplified, see \cite{Raidal:2017mfl,Chen:2018czv,Ali-Haimoud:2018dau,Desjacques:2018wuu,Ballesteros:2018swv} for the effects of clustering for Gaussian cases. Also, \cite{Ali-Haimoud:2017rtz, Raidal:2018bbj} thoroughly revisited estimates of the PBH merger rate.} 
and a monochromatic PBH mass function, the probability that the coalescence occurs in the time interval $(t, t + dt)$ can be estimated as $dP_t\propto T^{-3/37}dt$ for $t<t_c$ and $dP_t\propto T^{-3/8}t_c^{29/56}dt$ for $t>t_c$, with $t_c,T\propto M^{-5/3}$. Hence, the event rate is $P_t \propto M^{-32/37}$ or $P_t \propto M^{-26/21}$. This shows that, for more massive PBH binaries, to get a reasonable event rate (say, $1 \mathrm{Gpc}^{-3}\mathrm{yr}^{-1}$), we will need a greater PBH fraction $f=\Omega_{\mathrm{PBH}}/\Omega_{\mathrm{DM}}$. However, several constraints have been put on large $f$ by Eridanus II \cite{Brandt:2016aco}, Planck \cite{Ali-Haimoud:2016mbv}, wide-binary disruption \cite{Quinn:2009dp} and millilensing of quasars \cite{Wilkinson:2001vv}. Hence, massive PBH mergers in such a simple setup are unlikely be observed by future space-based interferometers.

The inclusion of initial spatial clustering of PBHs changes the story. The possibilities of clustered PBHs are discussed in \cite{Khlopov:2014uda,Belotsky:2014kca, Belotsky:2018wph,Bringmann:2018mxj} and references therein. We will show that initial clustering can significantly enhance the detectability of gravitational waves from massive PBHs. This is because of an increased formation rate of PBH binaries inside the clusters.

\begin{figure}[htbp] \centering
    \includegraphics[width=0.45\textwidth]{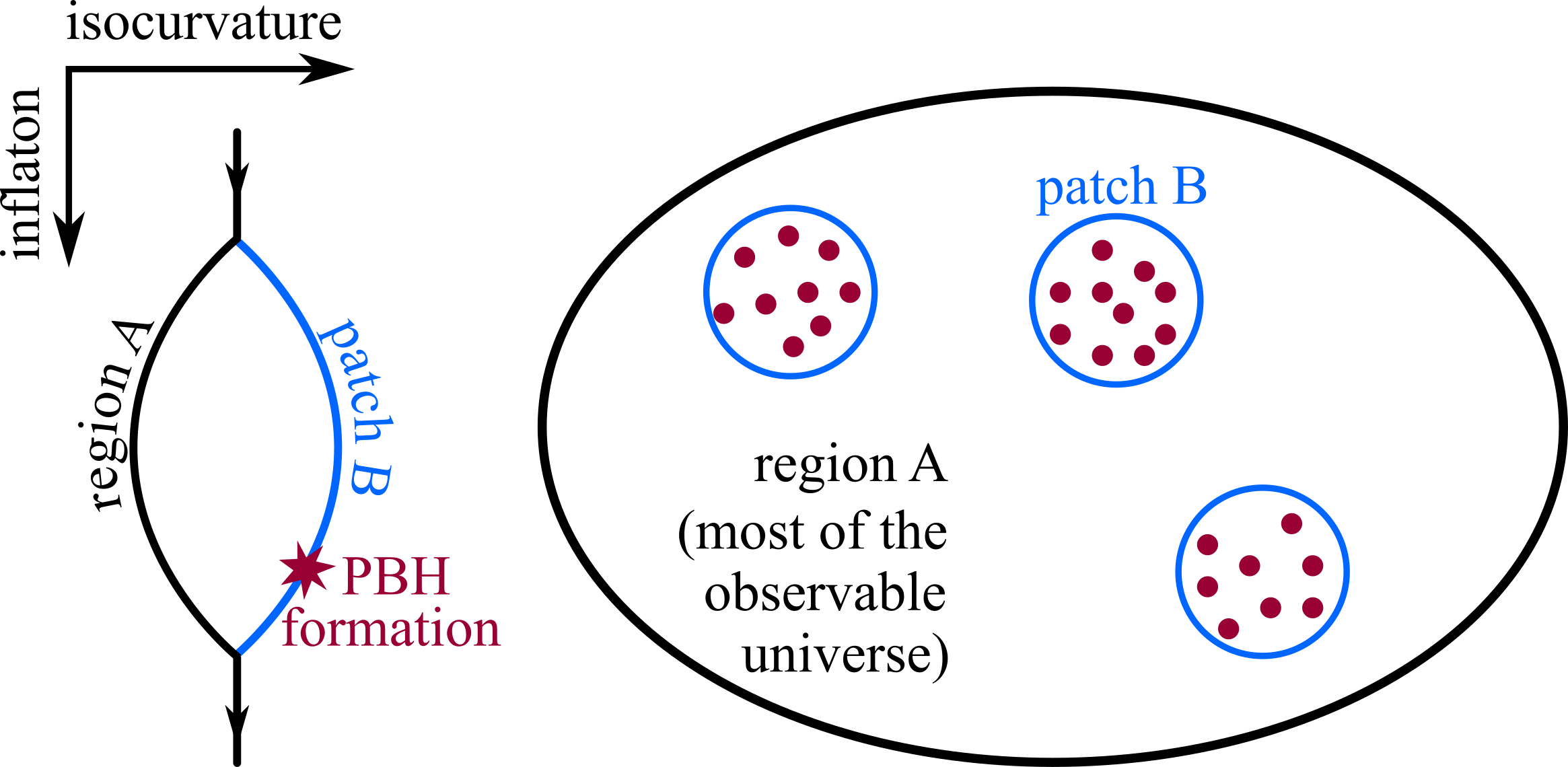}
    \caption{\label{fig:multistream}
        Illustration of clustered PBH production in multistream inflation. The left panel shows the field space trajectory during inflation involving two fields. The inflationary trajectory has bifurcated. In most cases, the inflationary trajectory is along the black line. However, with a small probability, the trajectory follows the blue line for a period during inflation, where fluctuations are enhanced around a particular scale and hence PBHs can be formed later. In our current universe, the black and blue trajectories manifest themselves as region A (most of the observable universe) and patches B with their combined volume fraction $\beta_1$.
    }
\end{figure}

To construct a simple model of clustering, we consider the scenario of multi-stream inflation \cite{Li:2009sp,Li:2009me,Wang:2010rs,Afshordi:2010wn} (see also \cite{Nakama:2016kfq,Gani:2017dgb} for a similar model). As illustrated in Fig.~\ref{fig:multistream}, the inflationary trajectory bifurcates at an encounter of a potential barrier in field space. Most of the observable universe (region A) follows one trajectory, while as rare events, a different trajectory with PBH formation
\footnote{For the PBH generation mechanisms, see, for example, \cite{Yokoyama:1995ex,Yokoyama:1998pt,Kawasaki:2016pql,Inomata:2016rbd}. Our discussion does not depend on the details of these PBH formation mechanisms, though for simplicity we assume the mass function is monochromatic. In addition to PBHs, ultracompact minihalos can also be formed in B patches after matter-radiation equality \cite{Nakama:2019htb}. However, they would not significantly affect the subsequent discussions, though they themselves would have interesting observational implications (see \cite{Nakama:2019htb} and references therein.)} leads to the
generation of  small bubbles (patch B) in the observable universe. Various inflaton potentials involving an inflection-type or plateau-type feature have been proposed to produce PBHs in single field inflation models, and one can assume any of such potential shapes along the trajectory B. Then the curvature perturbation is enhanced on small scales only in B patches, and when these perturbations reenter the horizon during radiation domination, a fraction of those would collapse to form PBHs. That is, only in B patches PBHs are formed during radiation domination. The volume fraction of B patches in our observable Universe can be controlled by the potential shape at the bifurcation point, and the radius of the B patches is determined by the inflaton field value at the bifurcation point. The mass of PBHs, formed only inside B patches, is controlled by the location of the inflection-type or plateau-type feature along the B trajectory, and the abundance of those PBHs can be controlled by the shape of such a feature, which determines the degree of the enhancement of the curvature perturbation. Multi-stream inflation may not be necessarily embedded in string theory, but its motivation may best be understood in the context of the string landscape, where inflation may have taken place in a multi-dimensional field space with a very complicated potential. Such a situation has been modeled by random potential in the literature. \cite{Liu:2015dda} showed that bifurcations of the inflaton trajectory can be common for such a random potential in high-dimensional field space.

We expect that our analysis should qualitatively apply for general models with significant initial clustering. However, the continuous variation of PBH densities in those models makes the analysis complicated. Thus, in this Letter, we focus on the clustering from multi-stream inflation, where there is no PBH in region A, and the PBH density is approximately  constant in B patches. We leave the detailed analysis of general clustering and continuously varying PBH density to  future work.

Let us denote the volume fraction of B patches by $\beta_1$ (the standard cases with small initial PBH clustering correspond to $\beta_1=1$), and the volume fraction of B patches collapsing to PBHs inside B patches by $\beta_2$. Then the initial abundance of PBHs is $\beta=\rho_{\mathrm{PBH}}/\rho_{\mathrm{rad}}=\beta_1\beta_2$, where the ratio should be evaluated at the moment of PBH formation. This is related to the present day density fraction of PBHs $f=\Omega_{\mathrm{PBH}}/\Omega_{\mathrm{DM}}$ via \cite{Nakama:2016gzw}
\begin{equation}
    \beta\sim 10^{-8}f\left(\frac{M}{30M_\odot}\right)^{1/2}.\label{beta}
\end{equation}
In order to be consistent with observations, we need $\beta\ll 1$. See \cite{Carr:2009jm} for limits on $\beta$ for different PBH masses. The smallness of $\beta_1$ controls the amount of initial clustering of PBHs for fixed $\beta$ or $f$. The local abundance $\beta_2$ is enhanced by $\beta_1^{-1}(>1)$ for fixed $\beta$, which leads to enhanced merger probability in B patches, as we show below.

There are several conditions restricting the parameter space of this model. Let us use $1/k_1$ to denote the comoving scale which exits the horizon at the time of bifurcation (and thus determines the size of B patches), and use $1/k_2$ to denote the comoving scale which exits the horizon at the time of PBH formation during inflation. The number $N$ of PBHs in each B patch is roughly $N=(k_2/k_1)^3\beta_2$. We assume $N>2^3$ or 
\begin{equation}
k_1< 2^{-1} \beta_2^{1/3}k_2.\label{number}
\end{equation}
Note that $\beta_2=\beta_1^{-1}\beta$ and $\beta$ is related to $f$ as Eq. (\ref{beta}) indicates, and also $k_2$ is related to the mass of PBHs as \cite{Nakama:2016gzw}
\begin{equation}
    k_2\simeq 7.5\times 10^5\mathrm{Mpc}^{-1}\left(\frac{M}{30M_\odot}\right)^{-1/2}. 
\end{equation}
Hence, the above inequality can be regarded as providing an upper limit on $k_1$, for each $(\beta_1,M,f)$. 

We also assume that gravitational wave experiments can cover a large number $N_B$ of B patches. The observable volume depends on the sensitivity of experiments. For simplicity, we fix the observable distance to be $k_{\mathrm{obs}}^{-1}= 1$Gpc. We express the above condition as
\begin{equation}
N_B=(k_1/k_{\mathrm{obs}})^3\beta_1> 2^3. \label{NB}
\end{equation}

In addition, we also focus on $\beta_2\ll1$, so that gravitationally-bound clusters of PBHs (PBH clusters) are not formed during radiation domination. This is to ensure the validity of the calculations of \cite{Sasaki:2016jop}, which we will use later. This condition can be quantified as follows. Let us denote the formation redshift of PBH clusters by $z_c$, which would be determined by $k_2$ and $\beta_2$. 
The calculations of \cite{Sasaki:2016jop} are only applicable before the formation of such PBH clusters. The formation redshift $z_c$ can be estimated as follows. PBH clusters would be formed approximately when B patches become locally-matter dominant during radiation domination, since the dynamical timescale of the would-be cluster then  becomes comparable to the Hubble timescale. Noting the ratio between the energy density of PBHs and that of radiation grows in proportion to the scale factor in B patches, we find $\beta_2 (z_*/z_c)=1$, or $z_c=\beta_2 z_*$, where $z_*$ is the redshift when PBHs are formed. We focus on the parameter region where $z_c<z_{\mathrm{eq}}$ is satisfied, so that the analysis of \cite{Sasaki:2016jop} can be used. 

Noting $\beta_2=\beta_1^{-1}\beta\simeq\beta_1^{-1}f(a_*/a_{\mathrm{eq}})$, the condition $z_c<z_{\mathrm{eq}}$ can also be rewritten as
\begin{equation}
    f<\beta_1. \label{beta1}
\end{equation}
This condition is roughly equivalent to the condition that PBHs comprise only a subdominant component of the dark matter in B patches.

With these constraints, let us now calculate the enhanced merger rate of PBHs with initial clustering. 

In \cite{Sasaki:2016jop}, the merger probability and event rate of PBH binaries were calculated for the standard cases with small initial clustering, \textit{i.e.} $\beta_1=1$. The merger probability crucially depends on $f=\Omega_{\mathrm{PBH}}/\Omega_{\mathrm{DM}}$. For more initial clustering cases with $\beta_1<1$, the merger probability is determined by the fraction $f_1$ of PBHs to the dark matter \textit{inside} B patches, that is, $f=\beta_1 f_1$. For the same $f$, $f_1$ is larger when $\beta_1<1$, so the merger probability is enhanced relative to the standard cases with $\beta_1=1$. Note that when the number density of black holes $n_{\mathrm{BH}}$ is multiplied by the merger probability per time $dP_c/dt$ calculated in \cite{Sasaki:2016jop}, in order to obtain the merger rate at some time $t$, the corresponding number density should be the average number density of black holes in the observed volume, instead of the local number density in B patches. And if the condition (\ref{NB}) is satisfied, then $f$ (instead of $f_1$) should be used for the calculation of $n_{\mathrm{BH}}$. That is, with clustering we have
\begin{equation}
    \mbox{(event rate)} = \frac{f\Omega_{\mathrm{DM}}\rho_c(t)}{M_{\mathrm{BH}}}\frac{dP_c}{dt}\bigg|_{f\rightarrow f_1},
\end{equation}
where $\rho_c$ denotes the critical density.

\begin{figure}[htbp]
    \begin{center}
      \includegraphics[clip,width=0.45\textwidth]{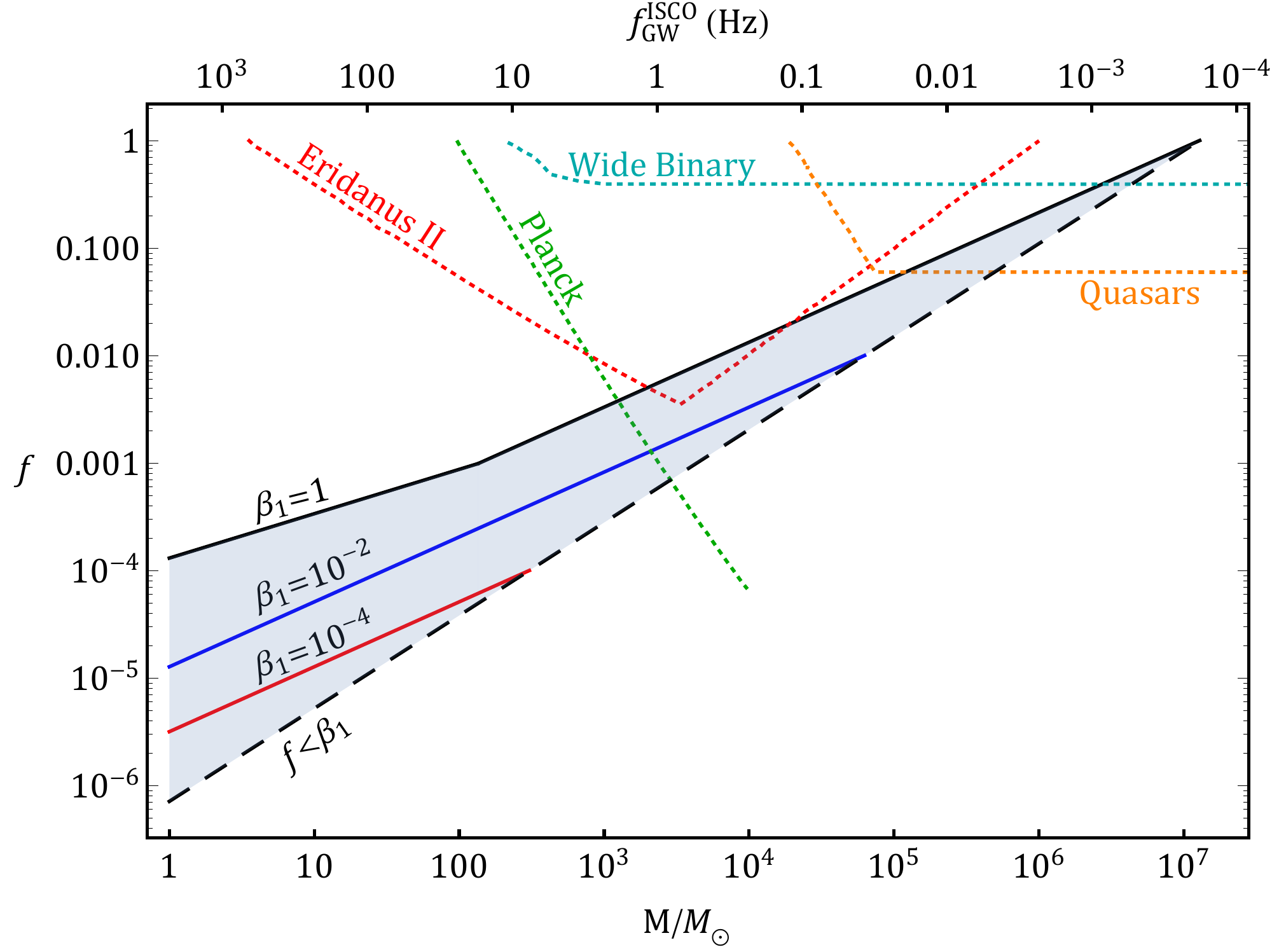}
      \caption{The fraction of PBHs $f=\Omega_{\mathrm{PBH}}/\Omega_{\mathrm{DM}}$ as a function of PBH mass $M$, needed to realize an event rate of $1\mathrm{Gpc}^{-3}\mathrm{yr}^{-1}$. We compare the cases of no clustering ($\beta_1=1$) with those of enhanced initial clustering ($\beta_1=10^{-2}, 10^{-4}$). We have taken $\Omega_{\mathrm{DM}} = 0.27$, $\mathrm{H}_{0} = 70\mathrm{km}\mathrm{Mpc}^{-1}\mathrm{s}^{-1}$, $z_{\mathrm{eq}} = 3000$ and $\mathrm{t}_{0} = 13.7\mathrm{Gyr}$. The space-based experiments are sensitive to the frequency range $10^{-3}\sim 1$Hz. Limits from Eridanus II \cite{Brandt:2016aco}, Plank \cite{Ali-Haimoud:2016mbv}, wide-binary disruption \cite{Quinn:2009dp} and millilensing of quasars \cite{Wilkinson:2001vv} are also plotted. Note that these limits are calculated assuming no initial clustering ($\beta_1=1$) and hence  cannot be directly applied to cases with initial clustering ($\beta_{1}<1$). We have selected a few bounds which are relatively simple for discussions on the impact of clustering, and see \cite{Sasaki:2018dmp} for a more complete collection of observational bounds.
      We also show the frequency of gravitational waves for each mass of merging PBHs along the upper horizontal axis, for which we use the frequency at the Innermost Stable Circular Orbit \cite{Centrella:2010mx}. }
      \label{fpbh}
    \end{center}
\end{figure}

For a fixed event rate, the fraction $f$ as a function of $M$ is plotted in Fig.~\ref{fpbh}. With clustering, smaller values of $f$ are needed to achieve the given event rate, as a result of the enhanced merger probability, controlled by the parameter $\beta_1<1$.

In Fig.~\ref{fpbh}, we have also included limits on massive PBHs \cite{Carr:2016drx}. It is important to note that these limits only apply for the case without initial clustering ($\beta_1=1$), in which case the parameter space for detecting PBH merger event from space-based interferometers is almost vanishing. However, they should not be interpreted as constraints for the cases of $\beta_1\ll 1$.

Three types of limits have been considered in Fig.~\ref{fpbh}. We briefly discuss the implication of clustering on these limits here: (1) Limits from the local density of PBHs within the local group, including Eridanus II \cite{Brandt:2016aco} and wide-binary disruption \cite{Quinn:2009dp}. When $\beta_1\ll 1$, the MWG halo and the Local Group are very unlikely to be in a PBH-rich region (patch B). So these limits do not constrain the case with strong initial clustering at all. (2) Millilensing of quasars \cite{Wilkinson:2001vv}. The validity of this limit will depend on the number and size of PBH-rich patches. If there are unlikely to be PBH-rich regions along all lines-of-sight between us and the quasars, the limit does not apply. Otherwise the quasar bound can apply. (3) CMB limits \cite{Ali-Haimoud:2016mbv} (see also \cite{Chen:2016pud, Poulin:2017bwe}). The effect of PBH accretion (which is an indirect limit depending on astrophysical assumptions) mainly affects the reionization history and thus modifies the temperature and polarization power spectra at low $\ell$. In addition, for larger mass PBHs (which is the focus of this Letter), the ionization effects shift to lower redshifts and thus affect even lower $\ell$. However, the initial clustering of PBHs is a high $\ell$ effect when  $\beta_1\ll 1$. Due to spatial inhomogeneity, the affected $\ell$ should be determined by the size of PBH-rich regions and the size of reionization bubbles, which correspond to smaller $\ell$. It remains interesting to work out the precise CMB limits considering initial clustering. We provide additional remarks on this issue later. Our results also imply that the constraints on PBHs based on the PBH merger rate shown in \cite{Ali-Haimoud:2017rtz, Kavanagh:2018ggo} are tentative, and they can get progressively stronger for increased clustering level.

We have assumed that the merger probability of PBHs up to the present moment $P_c(t_0)$, which can be calculated from the formulae in \cite{Sasaki:2016jop}, is sufficiently small so that the evolution of PBH population is negligible. This leads to an additional constraint on the model parameters. Let us require
\begin{equation}
    P_c(t_0)<0.1.\label{probability}
\end{equation}

Let us check whether the model parameters in Fig. \ref{fpbh} satisfy the conditions on the parameters. First let us consider the conditions which do not depend on $k_1$, that is, Eqs. (\ref{beta1}) and (\ref{probability}). Each solid line in Fig. \ref{fpbh} is terminated at some large $M$ so that these two conditions are satisfied along the lines. It turns out that Eq. (\ref{beta1}) is more restrictive than (\ref{probability}).

Next, let us consider the other conditions which depend on $k_1$, which are (\ref{number}) and (\ref{NB}). 
Of these, (\ref{number}) gives the upper bound on $k_1$, whereas (\ref{NB}) gives the lower bound on $k_1$, for fixed $f, M$ and $\beta_1$. 
Let us denote the upper bound on $k_1$, determined by (\ref{number}), as $k_M$, and the lower bound determined from (\ref{NB}) by $k_m$. 
If $k_m<k_M$, there exist values of $k_1$, which satisfy both conditions for fixed $f, M$ and $\beta_1$. It turned out that these conditions are much less restrictive than the condition given by Eq. (\ref{beta1}). 

To summarize, we have studied the detectability of PBH coalescence with initial clustering, with an emphasis on the more massive PBHs to be detected in the future by LISA-like experiments. 

In this work, we have adopted a simple model of initial clustering of PBHs, where the PBH-rich regions have sharp boundaries. But one may also generalize our discussions to other situations, such as other kinds of primordial non-Gaussianity causing initial clustering of PBHs \cite{Young:2015kda,Tada:2015noa}. Non-Gaussianity can also alleviate the fine-tuning problem associated with PBH formation \cite{Nakama:2018utx}. CMB distortion limits on massive PBHs \cite{Kohri:2014lza,Nakama:2017xvq} would also be avoided by clustering \cite{Nakama:2016kfq}, since it is suppressed by $\beta_1$ in our model. However, if $\beta_1$ is not so small, our scenario may also lead to CMB $\mu$ distortions observable by future high-sensitivity experiments for massive PBHs \cite{Nakama:2017xvq}. Our scenario may be tested by investigating spacial distributions of gravitational-wave events in details in future. Though we have restricted attention to the coalescence of PBH binaries at very low redshifts, following \cite{Sasaki:2016jop},  it would also be interesting to discuss high-redshift PBH mergers. It would also be worthwhile to investigate stochastic gravitational-wave backgrounds for PBHs of different masses with initial clustering. Depending on the model parameters, our scenario would cause inhomogeneous big bang nucleosynthesis, which might lead to interesting observational traces \cite{Carr:2018rid}. 

We have focused our attention on the parameter region $f<\beta_1$. This intersects a large part of the parameter regions of interest. However, we should note that this constraint is for theoretical simplicity instead of an observational bound. For $f>\beta_1$, in B patches PBHs dominate over other dark matter in energy density. Thus, the previous calculation of the coalescence time needs to be reconsidered. We will leave the detailed calculation to  future work, but add a few more remarks:

\textit{Evolution of PBH binaries}: We have assumed for simplicity that the PBH binary evolution is solely determined by the emission of GWs, but in principle it should also be affected by encounters between PBH binaries and, most importantly, single PBHs. Based on theoretical models and simulations, the authors of \cite{Raidal:2018bbj} claim that, when PBHs are the dominant component of dark matter, it is highly probable that PBH binaries are disrupted by other PBHs. However, when the PBH fraction is much smaller than unity, which holds true in the parameter space we restrict our analysis to, the impact of the disruption events seems negligible, as discussed there. See \cite{Vaskonen:2019jpv} for another recent discussion.

Even with the 1\%  fraction of 1-100 $M_\odot$ PBHs constrained by of PBH mergers rates, there is a substantial effect on dwarf galaxies. At this level of dynamical heating of ultradiffuse dwarf galaxies, the effects may actually be positive: rather than destruction of objects such as  Eridanus II. \cite{Brandt:2016aco}, one can actually induce a core cusp transition \cite{Boldrini:2019isx}. Moreover, some physical effects actually enhance coalescence, e.g. in dense clusters \cite{Petrovich:2017otm,Antonini:2018auk}. Also this may be true for galactic nuclei near a supermassive black hole \cite{Hamers:2018hxv}. Another example is the role of triples (which are very common for massive stars, though admittedly we don’t know too much about triple formation for PBHs. But if there are binaries there will be triples). See \cite{Antonini:2017ash} for mergers from triples. These have low spin, see \cite{Antonini:2017tgo,Rodriguez:2018jqu}.

\textit{Further clustering in the radiation-dominated era for $f>\beta_1$}: PBHs may further evolve and cluster in the early universe. Inside such PBH clusters, binary formation may be understood similarly to that in globular clusters \cite{Samsing:2017xmd,Samsing:2017oij,Samsing:2018ykz,Zevin:2018kzq}. Runaway tidal capture may also take place \cite{Stone:2016ryd}. GWs from hyperbolic encounters of PBHs can also yield interesting observational implications \cite{Garcia-Bellido:2017qal}. PBH clusters may also eventually form supermassive black holes, observed as high-redshift quasars (see \cite{Nakama:2016kfq} and references therein), whose origin still remains to be understood. 

\textit{Ultracompact minihalos and locally enhanced structure formation}: PBH-rich patches will be dark-matter overdense regions, which collapse much earlier than standard structure formation, which may also be called ultracompact minihalos \cite{Ricotti:2009bs}. In addition, Poisson fluctuations in the number density of B patches would also cause enhancement of structure formation, similarly to limits on massive PBHs due to the Poisson fluctuations in their number density \cite{Carr:2009jm}. These effects would be constrained by observations of the Lyman-$\alpha$ forest. If these constraints are too severe, one may consider some underdense regions to compensate the overdensities (which can also be realized in multi-stream inflation). Also, binary formation may be efficient in ultracompact minihalos. Moreover, in ultracompact minihalos, PBHs are likely to have relatively high velocities, so accretion effects may be suppressed, which would further weaken the CMB limits. Note also that small-scale structure formation may also be enhanced inside B patches due to Poisson fluctuations of PBHs, possibly leading to the formation of smaller ultracompact minihalos, before B patches themselves collapse. 

\bigskip\noindent\textit{Acknowledgements}: The research is supported in part by ECS Grant 26300316 and
GRF Grants 16301917 and 16304418 from the Research Grants Council of Hong Kong. 



\begin{thebibliography}{999}

  \bibitem{Hawking:1971ei} 
        S.~Hawking,
        ``Gravitationally collapsed objects of very low mass,''
        Mon.\ Not.\ Roy.\ Astron.\ Soc.\  {\bf 152}, 75 (1971).
  
  \bibitem{Dovich:1966no} 
        Y.~B.~Zel'dovich and I.~D.~Novikov,
        ``The Hypothesis of Cores Retarded during Expansion and the Hot Cosmological Model,''
         Astron.\ Zh.\   {\bf 43}, 758 (1966);
         Sov.\ Astron.\   {\bf 10}, 602 (1967).
  
  \bibitem{Carr:1974nx} 
        B.~J.~Carr and S.~W.~Hawking,
        ``Black holes in the early Universe,''
        Mon.\ Not.\ Roy.\ Astron.\ Soc.\  {\bf 168}, 399 (1974).

  \bibitem{Hawking:1974rv} 
        S.~W.~Hawking,
        ``Black hole explosions,''
        Nature {\bf 248}, 30 (1974).
  
  \bibitem{Carr:2009jm} 
        B.~J.~Carr, K.~Kohri, Y.~Sendouda and J.~Yokoyama,
        ``New cosmological constraints on primordial black holes,''
        Phys.\ Rev.\ D {\bf 81}, 104019 (2010)
        [arXiv:0912.5297 [astro-ph.CO]].
  
  \bibitem{Carr:2016drx} 
        B.~Carr, F.~Kuhnel and M.~Sandstad,
        ``Primordial Black Holes as Dark Matter,''
        Phys.\ Rev.\ D {\bf 94}, no. 8, 083504 (2016)
        [arXiv:1607.06077 [astro-ph.CO]].

  
  \bibitem{Sasaki:2018dmp} 
        M.~Sasaki, T.~Suyama, T.~Tanaka and S.~Yokoyama,
        ``Primordial black holes—perspectives in gravitational wave astronomy,''
        Class.\ Quant.\ Grav.\  {\bf 35}, no. 6, 063001 (2018)
        [arXiv:1801.05235 [astro-ph.CO]].
  
        
  \bibitem{Bird:2016dcv} 
        S.~Bird, I.~Cholis, J.~B.~Muñoz, Y.~Ali-Haïmoud, M.~Kamionkowski, E.~D.~Kovetz, A.~Raccanelli and A.~G.~Riess,
        ``Did LIGO detect dark matter?,''
        Phys.\ Rev.\ Lett.\  {\bf 116}, no. 20, 201301 (2016)
        [arXiv:1603.00464 [astro-ph.CO]].
  
  \bibitem{Clesse:2016vqa} 
        S.~Clesse and J.~García-Bellido,
        ``The clustering of massive Primordial Black Holes as Dark Matter: measuring their mass distribution with Advanced LIGO,''
        Phys.\ Dark Univ.\  {\bf 15}, 142 (2017)
        [arXiv:1603.05234 [astro-ph.CO]].
  
  \bibitem{Sasaki:2016jop} 
        M.~Sasaki, T.~Suyama, T.~Tanaka and S.~Yokoyama,
        ``Primordial Black Hole Scenario for the Gravitational-Wave Event GW150914,''
        Phys.\ Rev.\ Lett.\  {\bf 117}, no. 6, 061101 (2016)
        Erratum: [Phys.\ Rev.\ Lett.\  {\bf 121}, no. 5, 059901 (2018)]
        [arXiv:1603.08338 [astro-ph.CO]].
  
  \bibitem{Abbott:2016blz} 
        B.~P.~Abbott {\it et al.} [LIGO Scientific and Virgo Collaborations],
        ``Observation of Gravitational Waves from a Binary Black Hole Merger,''
        Phys.\ Rev.\ Lett.\  {\bf 116}, no. 6, 061102 (2016)
        [arXiv:1602.03837 [gr-qc]].
  
  \bibitem{Abramovici:1992ah} 
      A.~Abramovici {\it et al.},
      ``LIGO: The Laser interferometer gravitational wave observatory,''
      Science {\bf 256}, 325 (1992).
  
  \bibitem{TheVirgo:2014hva} 
      F.~Acernese {\it et al.} [VIRGO Collaboration],
      ``Advanced Virgo: a second-generation interferometric gravitational wave detector,''
      Class.\ Quant.\ Grav.\  {\bf 32}, no. 2, 024001 (2015)
      [arXiv:1408.3978 [gr-qc]].
  
  \bibitem{Somiya:2011np} 
      K.~Somiya [KAGRA Collaboration],
      ``Detector configuration of KAGRA: The Japanese cryogenic gravitational-wave detector,''
      Class.\ Quant.\ Grav.\  {\bf 29}, 124007 (2012)
      [arXiv:1111.7185 [gr-qc]].
  
  \bibitem{LISA}
      P. L. Bender, et al., LISA Pre-Phase A Report; Second Edition, MPQ 233 (1998).
  
  \bibitem{Kawamura:2006up} 
      S.~Kawamura {\it et al.},
      ``The Japanese space gravitational wave antenna DECIGO,''
      Class.\ Quant.\ Grav.\  {\bf 23}, S125 (2006).
  
  \bibitem{Hu:2017mde} 
      W.~R.~Hu and Y.~L.~Wu,
      ``The Taiji Program in Space for gravitational wave physics and the nature of gravity,''
      Natl.\ Sci.\ Rev.\  {\bf 4}, no. 5, 685 (2017).
  
  \bibitem{Luo:2015ght} 
      J.~Luo {\it et al.} [TianQin Collaboration],
      ``TianQin: a space-borne gravitational wave detector,''
      Class.\ Quant.\ Grav.\  {\bf 33}, no. 3, 035010 (2016)
      [arXiv:1512.02076 [astro-ph.IM]].
  
  \bibitem{Raidal:2017mfl} 
        M.~Raidal, V.~Vaskonen and H.~Veermäe,
        ``Gravitational Waves from Primordial Black Hole Mergers,''
        JCAP {\bf 1709}, 037 (2017)
        [arXiv:1707.01480 [astro-ph.CO]].
  
  \bibitem{Chen:2018czv} 
      Z.~C.~Chen and Q.~G.~Huang,
      ``Merger Rate Distribution of Primordial-Black-Hole Binaries,''
      Astrophys.\ J.\  {\bf 864}, no. 1, 61 (2018)
      [arXiv:1801.10327 [astro-ph.CO]].
  
  \bibitem{Ali-Haimoud:2018dau} 
        Y.~Ali-Haïmoud,
        ``Correlation Function of High-Threshold Regions and Application to the Initial Small-Scale Clustering of Primordial Black Holes,''
        Phys.\ Rev.\ Lett.\  {\bf 121}, no. 8, 081304 (2018)
        [arXiv:1805.05912 [astro-ph.CO]].
  
  \bibitem{Desjacques:2018wuu} 
        V.~Desjacques and A.~Riotto,
        ``Spatial clustering of primordial black holes,''
        Phys.\ Rev.\ D {\bf 98}, no. 12, 123533 (2018)
        [arXiv:1806.10414 [astro-ph.CO]].
  
  \bibitem{Ballesteros:2018swv} 
        G.~Ballesteros, P.~D.~Serpico and M.~Taoso,
        ``On the merger rate of primordial black holes: effects of nearest neighbours distribution and clustering,''
        JCAP {\bf 1810}, no. 10, 043 (2018)
        [arXiv:1807.02084 [astro-ph.CO]].
  
  \bibitem{Ali-Haimoud:2017rtz} 
        Y.~Ali-Haïmoud, E.~D.~Kovetz and M.~Kamionkowski,
        ``Merger rate of primordial black-hole binaries,''
        Phys.\ Rev.\ D {\bf 96}, no. 12, 123523 (2017)
        [arXiv:1709.06576 [astro-ph.CO]].
  
  \bibitem{Raidal:2018bbj} 
      M.~Raidal, C.~Spethmann, V.~Vaskonen and H.~Veermäe,
      ``Formation and Evolution of Primordial Black Hole Binaries in the Early Universe,''
      JCAP {\bf 1902}, 018 (2019)
      [arXiv:1812.01930 [astro-ph.CO]].
  
  \bibitem{Brandt:2016aco} 
        T.~D.~Brandt,
        ``Constraints on MACHO Dark Matter from Compact Stellar Systems in Ultra-Faint Dwarf Galaxies,''
        Astrophys.\ J.\  {\bf 824}, no. 2, L31 (2016)
        [arXiv:1605.03665 [astro-ph.GA]].
  
  \bibitem{Ali-Haimoud:2016mbv} 
        Y.~Ali-Haïmoud and M.~Kamionkowski,
        ``Cosmic microwave background limits on accreting primordial black holes,''
        Phys.\ Rev.\ D {\bf 95}, no. 4, 043534 (2017)
        [arXiv:1612.05644 [astro-ph.CO]].
  
  \bibitem{Quinn:2009dp} 
        D.~P.~Quinn, M.~I.~Wilkinson, M.~J.~Irwin, J.~Marshall, A.~Koch and V.~Belokurov,
        ``On the Reported Death of the MACHO Era,''
        MNRAS  {\bf 396}, 1, no. 1, L11-5 (2009)
        [arXiv:0903.1644 [astro-ph.GA]].
  
  \bibitem{Wilkinson:2001vv} 
        P.~N.~Wilkinson {\it et al.},
        ``Limits on the cosmological abundance of supermassive compact objects from a search for multiple imaging in compact radio sources,''
        Phys.\ Rev.\ Lett.\  {\bf 86}, 584 (2001)
        [astro-ph/0101328].
  
  \bibitem{Centrella:2010mx} 
        J.~Centrella, J.~G.~Baker, B.~J.~Kelly and J.~R.~van Meter,
        ``Black-hole binaries, gravitational waves, and numerical relativity,''
        Rev.\ Mod.\ Phys.\  {\bf 82}, 3069 (2010)
        [arXiv:1010.5260 [gr-qc]].

  \bibitem{Khlopov:2014uda} 
        M.~Y.~Khlopov and N.~A.~Chasnikov,
        ``Primordial Black Hole Clusters and their Evolution,''
        Bled Workshops Phys.\  {\bf 14}, no. 2, 107 (2013)
        [arXiv:1401.1643 [astro-ph.CO]].
  
  \bibitem{Belotsky:2014kca} 
        K.~M.~Belotsky {\it et al.},
        ``Signatures of primordial black hole dark matter,''
        Mod.\ Phys.\ Lett.\ A {\bf 29}, no. 37, 1440005 (2014)
        [arXiv:1410.0203 [astro-ph.CO]].
  
  \bibitem{Belotsky:2018wph} 
        K.~M.~Belotsky {\it et al.},
        ``Clusters of primordial black holes,''
        arXiv:1807.06590 [astro-ph.CO].
  
  \bibitem{Bringmann:2018mxj} 
        T.~Bringmann, P.~F.~Depta, V.~Domcke and K.~Schmidt-Hoberg,
        ``Strong constraints on clustered primordial black holes as dark matter,''
        arXiv:1808.05910 [astro-ph.CO].
  
  \bibitem{Li:2009sp} 
        M.~Li and Y.~Wang,
        ``Multi-Stream Inflation,''
        JCAP {\bf 0907}, 033 (2009)
        [arXiv:0903.2123 [hep-th]].
  
  \bibitem{Li:2009me} 
      S.~Li, Y.~Liu and Y.~S.~Piao,
      ``Inflation in Web,''
      Phys.\ Rev.\ D {\bf 80}, 123535 (2009)
      [arXiv:0906.3608 [hep-th]].
  
  \bibitem{Wang:2010rs} 
        Y.~Wang,
        ``Multi-Stream Inflation: Bifurcations and Recombinations in the Multiverse,''
        arXiv:1001.0008 [hep-th].
  
  \bibitem{Afshordi:2010wn} 
        N.~Afshordi, A.~Slosar and Y.~Wang,
        ``A Theory of a Spot,''
        JCAP {\bf 1101}, 019 (2011)
        [arXiv:1006.5021 [astro-ph.CO]].
  
  \bibitem{Nakama:2016kfq} 
        T.~Nakama, T.~Suyama and J.~Yokoyama,
        ``Supermassive black holes formed by direct collapse of inflationary perturbations,''
        Phys.\ Rev.\ D {\bf 94}, no. 10, 103522 (2016)
        [arXiv:1609.02245 [gr-qc]].

  \bibitem{Gani:2017dgb} 
        V.~A.~Gani, A.~A.~Kirillov and S.~G.~Rubin,
        ``Classical transitions with the topological number changing in the early Universe,''
        JCAP {\bf 1804}, no. 04, 042 (2018)
        [arXiv:1704.03688 [hep-th]].
  
  \bibitem{Yokoyama:1995ex} 
        J.~Yokoyama,
        ``Formation of MACHO primordial black holes in inflationary cosmology,''
        Astron.\ Astrophys.\  {\bf 318}, 673 (1997)
        [astro-ph/9509027].
  
  \bibitem{Yokoyama:1998pt} 
        J.~Yokoyama,
        ``Chaotic new inflation and formation of primordial black holes,''
        Phys.\ Rev.\ D {\bf 58}, 083510 (1998)
        [astro-ph/9802357].
  
  \bibitem{Kawasaki:2016pql} 
        M.~Kawasaki, A.~Kusenko, Y.~Tada and T.~T.~Yanagida,
        ``Primordial black holes as dark matter in supergravity inflation models,''
        Phys.\ Rev.\ D {\bf 94}, no. 8, 083523 (2016)
        [arXiv:1606.07631 [astro-ph.CO]].
  
  \bibitem{Inomata:2016rbd} 
        K.~Inomata, M.~Kawasaki, K.~Mukaida, Y.~Tada and T.~T.~Yanagida,
        ``Inflationary primordial black holes for the LIGO gravitational wave events and pulsar timing array experiments,''
        Phys.\ Rev.\ D {\bf 95}, no. 12, 123510 (2017)
        [arXiv:1611.06130 [astro-ph.CO]].
 
  \bibitem{Nakama:2019htb} 
        T.~Nakama, K.~Kohri and J.~Silk,
        ``Ultracompact minihalos associated with stellar-mass primordial black holes,''
        Phys.\ Rev.\ D {\bf 99}, no. 12, 123530 (2019)
        [arXiv:1905.04477 [astro-ph.CO]].

  \bibitem{Liu:2015dda} 
        J.~Liu, Y.~Wang and S.~Zhou,
        ``Nonuniqueness of classical inflationary trajectories on a high-dimensional landscape,''
        Phys.\ Rev.\ D {\bf 91}, no. 10, 103525 (2015)
        [arXiv:1501.06785 [hep-th]].
  
  \bibitem{Nakama:2016gzw} 
        T.~Nakama, J.~Silk and M.~Kamionkowski,
        ``Stochastic gravitational waves associated with the formation of primordial black holes,''
        Phys.\ Rev.\ D {\bf 95}, no. 4, 043511 (2017)
        [arXiv:1612.06264 [astro-ph.CO]].
  
  \bibitem{Chen:2016pud} 
      L.~Chen, Q.~G.~Huang and K.~Wang,
      ``Constraint on the abundance of primordial black holes in dark matter from Planck data,''
      JCAP {\bf 1612}, no. 12, 044 (2016)
      [arXiv:1608.02174 [astro-ph.CO]].
  
  \bibitem{Poulin:2017bwe} 
        V.~Poulin, P.~D.~Serpico, F.~Calore, S.~Clesse and K.~Kohri,
        ``CMB bounds on disk-accreting massive primordial black holes,''
        Phys.\ Rev.\ D {\bf 96}, no. 8, 083524 (2017)
        [arXiv:1707.04206 [astro-ph.CO]].
  
  \bibitem{Kavanagh:2018ggo} 
        B.~J.~Kavanagh, D.~Gaggero and G.~Bertone,
        ``Merger rate of a subdominant population of primordial black holes,''
        Phys.\ Rev.\ D {\bf 98}, no. 2, 023536 (2018)
        [arXiv:1805.09034 [astro-ph.CO]].

  \bibitem{Young:2015kda} 
        S.~Young and C.~T.~Byrnes,
        ``Signatures of non-gaussianity in the isocurvature modes of primordial black hole dark matter,''
        JCAP {\bf 1504}, no. 04, 034 (2015)
        [arXiv:1503.01505 [astro-ph.CO]].
  
  \bibitem{Tada:2015noa} 
        Y.~Tada and S.~Yokoyama,
        ``Primordial black holes as biased tracers,''
        Phys.\ Rev.\ D {\bf 91}, no. 12, 123534 (2015)
        [arXiv:1502.01124 [astro-ph.CO]].
  
  \bibitem{Nakama:2018utx} 
        T.~Nakama and Y.~Wang,
        ``Do we need fine-tuning to create primordial black holes?,''
        Phys.\ Rev.\ D {\bf 99}, no. 2, 023504 (2019)
        [arXiv:1811.01126 [astro-ph.CO]].
  
  \bibitem{Kohri:2014lza} 
        K.~Kohri, T.~Nakama and T.~Suyama,
        ``Testing scenarios of primordial black holes being the seeds of supermassive black holes by ultracompact minihalos and CMB $\mu$-distortions,''
        Phys.\ Rev.\ D {\bf 90}, no. 8, 083514 (2014)
        [arXiv:1405.5999 [astro-ph.CO]].

  \bibitem{Nakama:2017xvq} 
        T.~Nakama, B.~Carr and J.~Silk,
        ``Limits on primordial black holes from $\mu$ distortions in cosmic microwave background,''
        Phys.\ Rev.\ D {\bf 97}, no. 4, 043525 (2018)
        [arXiv:1710.06945 [astro-ph.CO]].
  
  \bibitem{Carr:2018rid} 
        B.~Carr and J.~Silk,
        ``Primordial Black Holes as Generators of Cosmic Structures,''
        Mon.\ Not.\ Roy.\ Astron.\ Soc.\  {\bf 478}, no. 3, 3756 (2018)
        [arXiv:1801.00672 [astro-ph.CO]].

  \bibitem{Vaskonen:2019jpv} 
        V.~Vaskonen and H.~Veermäe,
        ``A lower bound on the primordial black hole merger rate,''
        arXiv:1908.09752 [astro-ph.CO].

  \bibitem{Boldrini:2019isx} 
        P.~Boldrini, Y.~Miki, A.~Y.~Wagner, R.~Mohayaee, J.~Silk and A.~Arbey,
        ``Primordial black holes as dark matter: cusp-to-core transition in low-mass dwarf galaxies,''
        arXiv:1909.07395 [astro-ph.CO].
        Submitted to MNRAS.

  \bibitem{Petrovich:2017otm} 
        C.~Petrovich and F.~Antonini,
        ``Greatly enhanced merger rates of compact-object binaries in non-spherical nuclear star clusters,''
        Astrophys.\ J.\  {\bf 846}, no. 2, 146 (2017)
        [arXiv:1705.05848 [astro-ph.HE]].

  \bibitem{Antonini:2018auk} 
        F.~Antonini, M.~Gieles and A.~Gualandris,
        ``Black hole growth through hierarchical black hole mergers in dense star clusters: implications for gravitational wave detections,''
        arXiv:1811.03640 [astro-ph.HE].
  
  \bibitem{Hamers:2018hxv} 
        A.~S.~Hamers, B.~Bar-Or, C.~Petrovich and F.~Antonini,
        ``The impact of vector resonant relaxation on the evolution of binaries near a massive black hole: implications for gravitational wave sources,''
        Astrophys.\ J.\  {\bf 865}, no. 1, 2 (2018)
        [arXiv:1805.10313 [astro-ph.HE]].

  \bibitem{Antonini:2017ash} 
        F.~Antonini, S.~Toonen and A.~S.~Hamers,
        ``Binary black hole mergers from field triples: properties, rates and the impact of stellar evolution,''
        Astrophys.\ J.\  {\bf 841}, no. 2, 77 (2017)
        [arXiv:1703.06614 [astro-ph.GA]].

  \bibitem{Antonini:2017tgo} 
        F.~Antonini, C.~L.~Rodriguez, C.~Petrovich and C.~L.~Fischer,
        ``Precessional dynamics of black hole triples: binary mergers with near-zero effective spin,''
        Mon.\ Not.\ Roy.\ Astron.\ Soc.\  {\bf 480}, no. 1, L58 (2018)
        [arXiv:1711.07142 [astro-ph.HE]].

  \bibitem{Rodriguez:2018jqu} 
        C.~L.~Rodriguez and F.~Antonini,
        ``A Triple Origin for the Heavy and Low-Spin Binary Black Holes Detected by LIGO/Virgo,''
        Astrophys.\ J.\  {\bf 863}, no. 1, 7 (2018)
        [arXiv:1805.08212 [astro-ph.HE]].

  \bibitem{Samsing:2017xmd} 
        J.~Samsing,
        ``Eccentric Black Hole Mergers Forming in Globular Clusters,''
        Phys.\ Rev.\ D {\bf 97}, no. 10, 103014 (2018)
        [arXiv:1711.07452 [astro-ph.HE]].
  
  \bibitem{Samsing:2017oij} 
        J.~Samsing, A.~Askar and M.~Giersz,
        ``MOCCA-SURVEY Database. I. Eccentric Black Hole Mergers during Binary–Single Interactions in Globular Clusters,''
        Astrophys.\ J.\  {\bf 855}, no. 2, 124 (2018)
        [arXiv:1712.06186 [astro-ph.HE]].
  
  \bibitem{Samsing:2018ykz} 
        J.~Samsing, D.~J.~D'Orazio, A.~Askar and M.~Giersz,
        ``Black Hole Mergers from Globular Clusters Observable by LISA and LIGO: Results from post-Newtonian Binary-Single Scatterings,''
        arXiv:1802.08654 [astro-ph.HE].
  
  \bibitem{Zevin:2018kzq} 
        M.~Zevin, J.~Samsing, C.~Rodriguez, C.~J.~Haster and E.~Ramirez-Ruiz,
        ``Eccentric Black Hole Mergers in Dense Star Clusters: The Role of Binary-Binary Encounters,''
        Astrophys.\ J.\  {\bf 871}, 1 (2019)
        [arXiv:1810.00901 [astro-ph.HE]].
  
  \bibitem{Stone:2016ryd} 
        N.~C.~Stone, A.~H.~W.~Küpper and J.~P.~Ostriker,
        ``Formation of Massive Black Holes in Galactic Nuclei: Runaway Tidal Encounters,''
        Mon.\ Not.\ Roy.\ Astron.\ Soc.\  {\bf 467}, no. 4, 4180 (2017)
        [arXiv:1606.01909 [astro-ph.GA]].
  
  \bibitem{Garcia-Bellido:2017qal} 
        J.~Garcia-Bellido and S.~Nesseris,
        ``Gravitational wave bursts from Primordial Black Hole hyperbolic encounters,''
        Phys.\ Dark Univ.\  {\bf 18}, 123 (2017)
        [arXiv:1706.02111 [astro-ph.CO]].
  
  \bibitem{Ricotti:2009bs} 
        M.~Ricotti and A.~Gould,
        ``A New Probe of Dark Matter and High-Energy Universe Using Microlensing,''
        Astrophys.\ J.\  {\bf 707}, 979 (2009)
        [arXiv:0908.0735 [astro-ph.CO]].
  
  \end{thebibliography}
\end{document}